\documentclass[twocolumn]{revtex4}
\usepackage[utf8]{inputenc}
\usepackage{graphicx}

\begin{document}

\title{Mean-field theory of the electromagnon resonance}

\author{Pavel A. Andreev}
\email{andreevpa@physics.msu.ru}
\affiliation{Department of General Physics, Faculty of physics, Lomonosov Moscow State University, Moscow, Russian Federation, 119991.}
\author{Mariya Iv. Trukhanova}
\email{trukhanova@physics.msu.ru}
\affiliation{Faculty of physics, Lomonosov Moscow State University, Moscow, Russian Federation, 119991.}
\affiliation{Russian Academy of Sciences, Nuclear Safety Institute (IBRAE), B. Tulskaya 52, Moscow, Russian Federation, 115191.}

\date{\today}

\begin{abstract}
We present the analytical theory of the electromagnon resonance for the multiferroics of spin origin.
We consider the spin density evolution under the influence of magnetoelectric coupling in the presence of the electromagnetic wave.
The dielectric permeability is found for the eigen-wave perturbations accompanied by perturbations of the electric field.
The imaginary part of the dielectric permeability is found as the function of the applied electric field frequency,
while the frequency of the eigen-waves is found from the dispersion equation as the function of the wave vector and parameters of the system.
The result shows the existence of two peaks.
One sharp peak is associated with the magnon resonance,
while the second wide peak at the approximately four times smaller frequency is interpreted as the electromagnon resonance in accordance with existing experimental data.
\end{abstract}


\maketitle



Direct influence of the electric field on the magnons/spin waves leads to the concept of the electromagnons,
where the action of the electric part of the electromagnetic wave on the spin waves happens \cite{Mostovoy npj 24}.
This effect is associated with the multiferroic materials,
where interference between magnetic and dielectric properties is observable.

The long-wavelength limit of the spin waves can be considered for the uniform medium,
where the equilibrium spin density does not depend on the space coordinates.
However, the appearance of electric polarization of the spin origin requires some mediation of spin density.
Formally, it is possible to include the effects responsible for polarization in the Landau--Lifshitz--Gilbert equation
and consider the spin wave perturbations of the uniform equilibrium state under the influence of the electromagnetic wave.
In this scenario, the periodic change of the electric field in the electromagnetic wave may cause some dielectric response.
This leads to the analysis of electromagnons, obtaining their dispersion dependence
and corresponding dielectric permeability.
Another scenario for the electromagnon analysis is possible
at the application of the equilibrium external constant electric field
acting on the realistic equilibrium spin density depending on the space coordinate in the well-known regime,
where the electric polarization of spin origin exists.
There are possible regimes of modeling the electromagnons.
However, it is necessary to identify the macroscopic parameters and their behavior for the comparison of the theory and the experiment.
One of the methods of detection of the electromagnons is the observation of the additional wide peak in the imaginary part of the dielectric permeability
\cite{Pimenov NP 06}, \cite{Pimenov PRB 06}, \cite{Pimenov PRL 09}, \cite{Pimenov JP CM 08}, \cite{ShuvaevPimenov EPJB 11}, \cite{Aupiais npj QM 18}.

Electromagnons are the collective excitations existing in magnetically ordered materials due to magneto-electric coupling.
Experimentally, electromagnons demonstrate themselves in the spectrum of the imaginary part of the dielectric permeability $\varepsilon$,
where the additional
low and wide
resonance is observed \cite{Pimenov PRB 06}
(see also \cite{Pimenov NP 06},
where the electromagnon resonance is demonstrated with no comparison to the magnon resonance).
It appears at frequencies that are 2-5 times smaller
than
the frequency of the magnon resonance,
which also reveals itself in the dielectric permeability.
It is also demonstrated in Ref. \cite{Pimenov PRB 06}
that the presence of the magnetic field of the order of $2$ T reduces electromagnon resonance,
so it almost disappears.

The appearance of the dielectric permeability in the system of magnetic moments is related to the magnetoelectric effect,
forming the electric polarization related to some spin formations due to the spin-orbit coupling.
It can form both static and dynamic polarizations.
Regarding the electromagnons,
we need to focus on the dynamic effects.
It allows us to neglect properties of the static polarization
(including the regime of zero static polarization).

The Dzyaloshinskii-Moriya interaction breaks the symmetry of the dispersion dependence of the spin waves,
introducing the dependence on the projection of the wave vector
\cite{Wang PRL 15}, \cite{Moon PRB 13}, \cite{Zakeri PRL 10}. 
Hence, the Dzyaloshinskii-Moriya interaction does not cause any instability
that leads to the reorganization of the equilibrium state
(for the linear regime on the small amplitude perturbations).
Similar behavior of the dispersion dependence happens at the consideration of the magnetoelectric coupling \cite{ZvezdinMukhin JETP L 09}.
Effective axis of asymmetry appears in this regime due to the action of the electric field.
There is some similarity in the structure of the spin torque of the Dzyaloshinskii-Moriya interaction
and the magnetoelectric coupling, but the last one gives the explicit contribution of the electric field.
Therefore, we do not expect
that the Dzyaloshinskii-Moriya interaction gives the major contribution to the appearance of the electromagnon,
so it is not considered in the model applied in this paper.

In order to understand the nature of the electromagnons,
let us discuss the nature of the magnetoelectric coupling,
which is related to the spin-current model \cite{Sergienko PRL 06}, \cite{Katsura PRL 05}, \cite{Sergienko PRB 06}
(see also review articles \cite{Tokura RPP 14}, \cite{Dong AinP 15}).
Further generalization and analysis of the spin-current model presented in Refs.
\cite{AndreevTrukh PS 24}, \cite{AndreevTrukh JETP 24},
where the analytical and numerical estimations of the constant
with the clarification of the mechanism of the polarization formation are given.
The specification of the interaction responsible for the magnetoelectric coupling is also given in Refs.
\cite{AndreevTrukh PS 24}, \cite{AndreevTrukh JETP 24}.
It is shown
that the simultaneous action of the spin-orbit interaction and the Coulomb exchange interaction
leads to the formation of polarization related to noncollinear spins.
The Heisenberg exchange interaction leads to the formation of the parallel order of spins.
If other interactions
create the condition for the noncollinear spin order,
then the Heisenberg exchange interaction together with the
spin-orbit interaction can form the polarization on this background
as the second level of this hierarchy \cite{AndreevTrukh PS 24}.

The electromagnons are associated with the response of electric polarization to the action of the electric field of the electromagnetic wave,
while electric polarization appears in systems of magnetic moments or spins.
There are several regimes of the magneto-electric coupling,
but the regime most actively studied in multiferroics
shows the appearance of the electric dipole moments related to the noncollinear order of spins
\cite{Tokura RPP 14}, \cite{Khomskii JETP 21}
$\textbf{d}_{ij}= \alpha_{ij}[\textbf{r}_{ij}\times[\textbf{s}_{i}\times\textbf{s}_{j}]]$,
where $\alpha_{ij}$ is the function of the interparticle distance,
which can be considered as a constant in the approximation of particles located at lattice points.
This leads to the macroscopic polarization \cite{Sparavigna PRB 94}, \cite{Mostovoy PRL 06}
$\textbf{P}= \sigma
[\textbf{S}(\nabla\cdot \textbf{S})-(\textbf{S}\cdot\nabla)\textbf{S}]$,
where the constant $\sigma$ is a macroscopic remnant of the microscopic function $\alpha_{ij}$
(possible generalizations of the magnetoelectric coupling are discussed in \cite{Solovyev PRL 21}).
It is demonstrated that the constants $\alpha_{ij}$ and $\sigma$ are related
to the simultaneous action of the Heisenberg exchange interaction and the spin-orbit interaction \cite{AndreevTrukh PS 24}.
Hence, the constant $\sigma$ is basically proportional to the exchange integral of the symmetric Heisenberg Hamiltonian.
The microscopically-macroscopic quantum hydrodynamic model developed in Ref. \cite{AndreevTrukh PS 24}
shows the dynamical nature of the spin-current model
\cite{Sergienko PRL 06}, \cite{Katsura PRL 05}, \cite{Sergienko PRB 06}.
Hence, the equation $\textbf{P}=
\sigma
[\textbf{S}(\nabla\cdot \textbf{S})-(\textbf{S}\cdot\nabla)\textbf{S}]$
can be considered both in the static and dynamic regimes.
Hence, the electric polarization evolution equation can be obtained on this background
\cite{AndreevTrukh PS 24}, \cite{AndreevTrukh JETP 24}, \cite{AndreevTrukh EPJ B 24}.
To some extent it corresponds to the concept suggested in Ref. \cite{JuraschekBalatsky PRM 17},
where some dynamical aspects of multiferroicity are discussed.

In order to identify the physical mechanisms of the electromagnon resonance
one needs to consider the most simple model structure.
While the experimentally studied multiferroic materials have the noncollinear equilibrium spin structure of the ferromagnetic or antiferromagnetic type,
we focus on the dynamical nature of the electromagnons.
Therefore, we consider the simple equilibrium of collinear spins,
so the equilibrium polarization is equal to zero as well.
Polarization in this regime appears only due to the dynamic of the magnetic moments,
which creates the sinusoidal noncollinear order on the macroscopic scale corresponding to the long-wavelength spin-wave excitations.
The described stationery state perturbed by the small amplitude spin-wave and electromagnetic excitations propagating in the medium
can be described in the following form:
$\textbf{S}_{0}=S_{0}\textbf{e}_{z}$ for the spin density,
and
$\textbf{E}_{0}=\epsilon_{0}\textbf{e}_{y}\cos(\omega_{0} t-k_{0}x)$ for the periodic electric field of the externally applied electromagnetic wave
with the following dispersion dependence $\omega_{0}=k_{0}c$,
which can be approximately applied far from the spin wave dispersion dependence
(the speed of light $v_{l}$ in the dielectric medium at the nonresonance propagation of the wave can be more accurate,
but we neglect this difference in our analysis).
Here, we consider the propagation of the linearly polarized wave in the direction perpendicular to the equilibrium spin direction.
Moreover, the direction of the electric field in the external wave is also perpendicular to the equilibrium spin direction.
We focus on the small-amplitude perturbations propagating in the same direction as the external electromagnetic wave.

The mean-field dynamic of the described system is modeled within
the Landau--Lifshitz--Gilbert equation
containing the magneto-electric coupling corresponding to the electric dipole moment shown above:
$$\partial_{t}\textbf{S}=
A[\textbf{S}\times\triangle\textbf{S}]
+\kappa [\textbf{S}\times S_{z}\textbf{e}_{z}] $$
$$-\sigma
\biggl[ [\textbf{E}\times \nabla] S^{2}
-2(\textbf{S}\cdot[\textbf{E}\times\nabla]) \textbf{S}$$
\begin{equation}\label{MFMemf s evolution MAIN TEXT}
-S^{2}(\nabla\times\textbf{E})
+\textbf{S}(\textbf{S}\cdot [\nabla\times\textbf{E}])
\biggr]
+a[\textbf{S}\times\partial_{t}\textbf{S}]. \end{equation}
The presented magneto-electric coupling corresponds to \cite{Risinggard SR 16},
where it is demonstrated for the simplified form of the constant electric field.
The accounting of the eigen-electromagnetic waves in the medium requires consideration of the Maxwell's equations along with
the Landau--Lifshitz--Gilbert equation (\ref{MFMemf s evolution MAIN TEXT}).
Each function in equation (\ref{MFMemf s evolution MAIN TEXT}) describing the evolution of the perturbations in the system is presented as the sum of
the stationary part and the perturbation
$\textbf{S}=\textbf{S}_{0}+\delta \textbf{S}$,
$\textbf{B}=\textbf{B}_{0}+\delta \textbf{B}$,
and
$\textbf{E}=\textbf{E}_{0}+\delta \textbf{E}$.
Here, the parameters $\textbf{S}_{0}$ and $\textbf{E}_{0}$ are described above,
while the magnetic field is assumed to be equal to zero
$\textbf{B}_{0}=0$.
It's a model simplified description,
since there is the magnetic field of the external wave,
but it is not included in the model and its influence is left for further consideration.

It leads to the linearized form of the Landau--Lifshitz--Gilbert equation
$$\partial_{t}\delta\textbf{S}=
A[\textbf{S}_{0}\times\triangle\delta\textbf{S}]
-\kappa [\textbf{S}_{0}\times \delta\textbf{S}]
+a[\textbf{S}\times\partial_{t}\delta\textbf{S}]$$
\begin{equation}\label{MFMemf s evolution lin MAIN TEXT}
+\sigma
\biggl[ S_{0}^{2}(\textbf{e}_{x}\times\partial_{x}\delta\textbf{E})
-2S_{0} \epsilon_{0}\cos(\omega_{0} t-k_{0}x)\partial_{x}\delta\textbf{S}\biggr]
, \end{equation}
which is coupled to the linearized Maxwell's equations.

The presence of the periodic coefficient in the last term of equation (\ref{MFMemf s evolution lin MAIN TEXT}) leads to the account
of the harmonics of the frequency and wave vector of the external wave,
in addition to the frequency of the eigen-waves.
We include the first harmonics and their amplitudes
$\delta\textbf{S}=\sum_{n=-1}^{+1}S_{n}^{\alpha}e^{-\imath\omega t+\imath kx-\imath\omega_{0}nt+\imath k_{0}nx}$
due to the small value of the coefficient $\sigma$ and the amplitude of the external wave $\epsilon_{0}$
(both of them combine in the characteristic frequency,
which should be small in comparison with the characteristic frequency of the electromagnetic and spin waves existed in the system).

\begin{figure}
\includegraphics[width=8cm,angle=0]{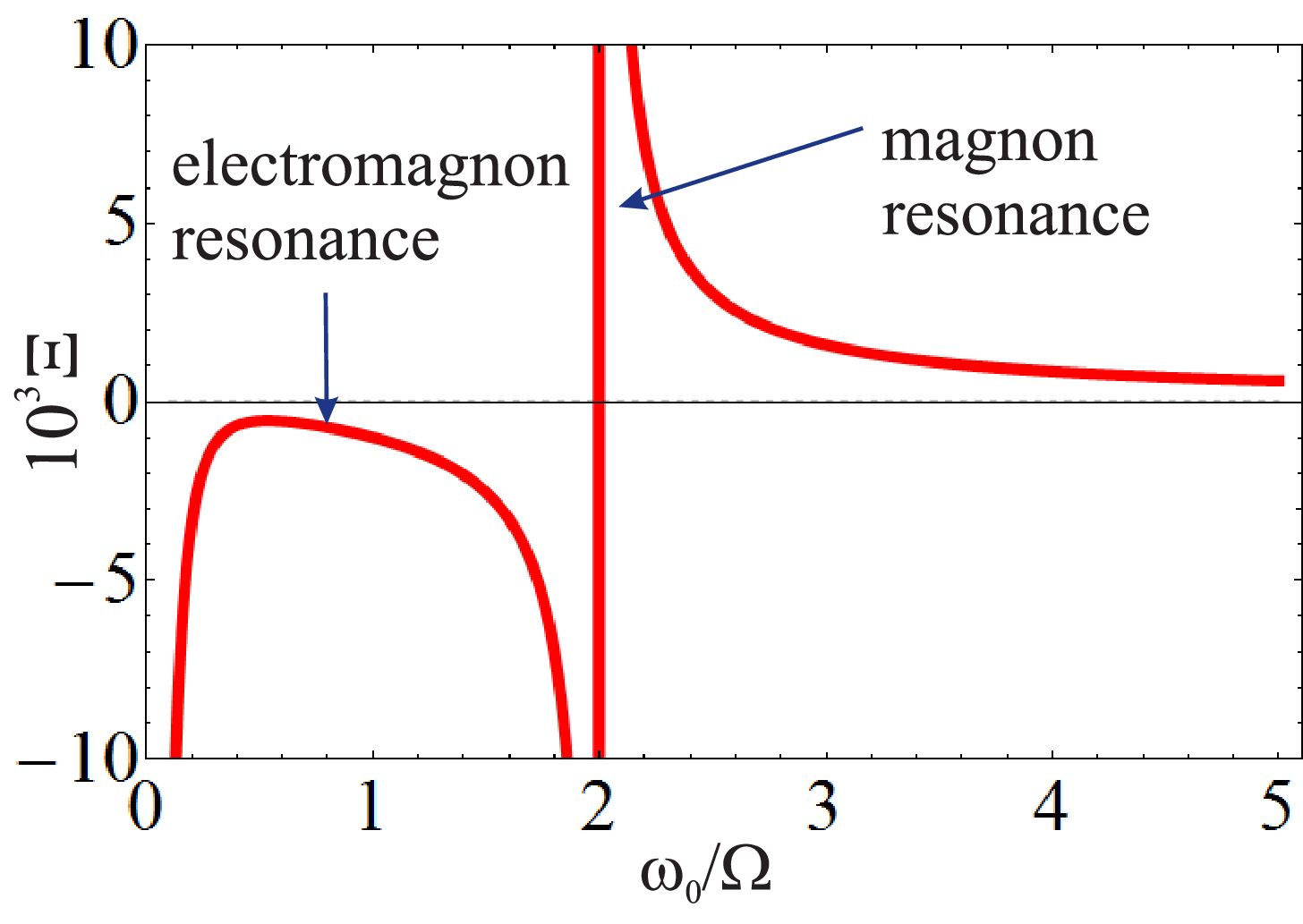}
\caption{\label{MFMextremumP Fig 02}
The resonance part of the imaginary part of
the dielectric permeability
$Im\varepsilon_{zz}=\kappa_{0}[(1+\tilde{a}^{2})^{-1}|\tilde{a}|
+\frac{(sk)^{2}}{\Omega^{2}}(1+\tilde{a}^{2})^{3}\Xi]$
is shown as the function of the dimensionless frequency of the external electric field,
which is proportional to the dimensionless parameter $\Xi$ being the function of the dimensionless frequency of the external field $\omega_{0}/\Omega$.
Explicitly, function $\Xi$ is presented in the
\emph{Supplementary materials}
(see Eq. (\ref{MFMemf Xi explicit}))
Coefficient $\kappa_{0}$ is presented by equation (\ref{MFMemf kappa 0}).
Arrows show the position of the low-frequency
electromagnon resonance and
the high-frequency
magnon resonance.
The figure is made for $\tilde{a}=-0.05$.}
\end{figure}

The anisotropy energy gives the characteristic frequency of the system $\Omega_{an}=\kappa S_{0}$,
which is of the order of $10^{10}\div10^{12}$ s$^{-1}$.
The contribution of the exchange energy modifies the frequency,
the long-wavelength limit can be considered up to wave vectors of order of $10^{5}\div10^{6}$ cm$^{-1}$.
While the electromagnetic wave with the frequency of the order of $10^{10}$ s$^{-1}$ has the small wave vectors of order of $1\div10$ cm$^{-1}$.
In reverse,
the electromagnetic wave with the wave vectors in the range $10^{3}\div10^{6}$ cm$^{-1}$ has the frequency several orders larger
than $\Omega_{an}$.
It shows that the spin eigenwaves and the electromagnetic eigenwaves exist on different scales of the frequencies at the fixed wave vector.
It allows us to consider the "low-frequency" regime $\omega\ll kc$,
where we exclude the electromagnetic eigen-waves from our analysis.
It explicitly shows in the simplification of the dispersion equation.
The described difference in scales gives the second simplification of the model.
Since we are interested in the frequency of the external electromagnetic wave $\omega_{0}$ comparable to or smaller
than $\Omega_{an}$.
It corresponds to the small wave vectors $k_{0}$ in comparison with the wave vectors of the spin eigen-waves $k$.
Hence, $k_{0}$ can be neglected in the spin perturbations and the dispersion equation.

Perturbations of the polarization are proportional to the perturbations of the spin density
$\delta P_{z,n=0}
=\sigma S_{0}\imath k\delta S_{x,0}/\gamma\mu$,
while the perturbations of the spin density as the function of the electric field
contain the magneto-electric coupling constant $\delta S_{x,0}\sim\sigma\delta E_{z,n=0}$.
In the lowest order on the relativistic effects $\delta S_{x,0}\sim 1/(\omega^{2}-\Omega_{0}^{2})$,
however, we find no resonance here,
since $\omega^{2}-\Omega_{0}^{2}$ can be replaced by the combination of terms of relativistic nature due to the dispersion equation
$\omega^{2}-\Omega_{0}^{2}\sim\sigma$.
Altogether, it shows that the main term in the polarization is proportional to the first degree of the magneto-electric coupling constant $\sigma$.

The main goal of this paper is the derivation of the imaginary part of
the dielectric permeability,
where
the dielectric permeability tensor,
that is defined as the relation between
complex amplitudes of the perturbations of the polarization and the electric field:
$\delta P_{z,n=0} \equiv \kappa_{zz} \delta E_{z,n=0}
\equiv (\varepsilon_{zz}-1) \delta E_{z,n=0}/4\pi$.

The final result for the complex dielectric susceptibility is found in the following form
(main steps of the derivation are discussed in
\emph{Supplementary materials})
$$\kappa_{zz}
= \sigma S_{0}\frac{k^{2}c}{4\pi\gamma^{2}\mu\omega}$$
\begin{equation}\label{MFMemf kappa zz fin MT}
\times
\frac{\biggl\{\Omega_{0}\mathcal{B}^{2}\mathcal{D}^{2}
+(sk)^2 \biggl[\mathcal{B}_{0}^{2}\Omega_{-}
+\mathcal{D}_{0}^{2}\Omega_{+}\biggr]\biggr\}}
{ \biggl\{
\omega
\mathcal{B}^{2}\mathcal{D}^{2}
+(sk)^2[\mathcal{B}_{0}^{2} (\omega-\omega_{0})
-\mathcal{D}_{0}^{2}(\omega+\omega_{0})] \biggr\}},
\end{equation}
where
we use $s\equiv \sigma S_{0}\epsilon_{0}$,
$\mathcal{B}_{0}^{2}\equiv (\omega+\omega_{0})^{2}-\Omega_{+}^{2}$,
and
$\mathcal{D}_{0}^{2}\equiv (\omega-\omega_{0})^{2}-\Omega_{-}^{2}$,
\emph{and}
$\mathcal{B}^{2}\equiv (\omega+\omega_{0})^{2}j-\Omega_{+}^{2}$,
and
$\mathcal{D}^{2}\equiv (\omega-\omega_{0})^{2}l-\Omega_{-}^{2}$,
\emph{and}
$j\approx l\approx 1+\alpha$,
$\alpha\equiv 4\pi\gamma\sigma S_{0}^{2}/c$,
\emph{and}
$\Omega_{0}\equiv (\kappa+\imath \omega a+Ak^{2})S_{0}$,
$\Omega_{+}\equiv (\kappa+\imath(\omega+\omega_{0})a+A(k+k_{0})^{2})S_{0}$,
and
$\Omega_{-}\equiv (\kappa+\imath(\omega-\omega_{0})a+A(k-k_{0})^{2})S_{0}$,
while $k_{0}$ can be neglected in compare with $k$.

The main part of the dielectric susceptibility
\begin{equation}\label{MFMemf kappa zz main 0 on alpha MT}
\kappa_{zz,main}
=\kappa_{0}\frac{\Omega}{\omega}\frac{\Omega_{0}}{\omega}
\approx\kappa_{0}\frac{\Omega}{\Omega_{0}}=\frac{\kappa_{0}}{1+\imath \tilde{a}}=\kappa_{0}\frac{1-\imath \tilde{a}}{1+\tilde{a}^{2}}
, \end{equation}
where we use the nonrelativistic part of the dispersion dependence $\omega=\Omega_{0}$
(or $\omega=\nu\Omega(1+\imath \tilde{a})$ with $\nu=1/(1+\tilde{a}^{2})$),
$\Omega\equiv(\kappa+Ak^{2})S_{0}$
\emph{and}
\begin{equation}\label{MFMemf kappa 0}
\kappa_{0}=\frac{\sigma\hbar S_{0}k^2 c}{4\pi\mu\Omega}.
\end{equation}
The relativistic part of the dispersion dependence can give small non-resonance dependence
of $\kappa_{0}\Omega\Omega_{0}/\omega^{2}$ on the frequency of the applied electric field $\omega_{0}$,
which is neglected,
since we are mostly interested in the resonance dependence on frequency $\omega_{0}$.
The equation (\ref{MFMemf kappa zz main 0 on alpha MT}) leads to the positive value of
the imaginary part of dielectric susceptibility $\kappa_{zz,main}=\nu\kappa_{0}\mid\tilde{a}\mid$ ($\tilde{a}<0$),
which describes the damping of the applied electric field in the medium.

Consideration of the first resonance relativistic correction to $\kappa_{zz,main}$ (\ref{MFMemf kappa zz main 0 on alpha MT}),
being proportional to $(sk)^{2}$,
following from equation (\ref{MFMemf kappa zz fin MT}),
we obtain the dielectric permeability
$\varepsilon_{zz}=Re\varepsilon_{zz}+\imath Im\varepsilon_{zz}$,
with the imaginary part of the following structure
$Im\varepsilon_{zz}=Im\kappa_{zz}=\kappa_{0}[(1+\tilde{a}^{2})^{-1}|\tilde{a}|
+\frac{(sk)^{2}}{\Omega^{2}}(1+\tilde{a}^{2})^{3}\Xi]$
(function $\Xi$ is demonstrated in Fig. 1).
The resonance dependence on $\omega_{0}$ is located in $\Xi=\Xi(\omega_{0})$,
but the rather huge explicit expression for function $\Xi(\omega_{0})$
is placed in the
\emph{Supplementary materials}.
The dependence of function $\Xi(\omega_{0})$ on the frequency of the external electric field is demonstrated
in Fig. \ref{MFMextremumP Fig 02}.
The general structure of $Im\varepsilon_{zz}\sim \Xi(\omega_{0})$ shows agreement with the experimentally observed results
(\cite{Pimenov NP 06},
\cite{Pimenov JP CM 08}
(see Figs. 3, 7 and 10),
\cite{ShuvaevPimenov EPJB 11}
(see Fig. 2)).
Ref. \cite{Pimenov NP 06} presents the result of new phenomenon observation
expressing itself in the existence of the additional wide peak,
while Ref. \cite{ShuvaevPimenov EPJB 11} gives additional information
about the relative position and the comparison of widths of peaks for the electromagnon resonance and the magnon resonance in the dielectric permeability.
In both cases of experiment and our results,
we see the wide peak in the low-frequency range existing in addition to the sharp peak corresponding to the magnon resonance.

In Ref. \cite{Holbein PRB 23}
the magnon dispersion
in the incommensurate cycloid phase is analyzed,
and it is found as a complex structure both
in the regime of neglecting Tb moments
and the regime of accounting for Tb moments.
The spiral magnetic arrangement shows the significant difference
between modes polarized perpendicular to the spiral plane
and modes polarized within this plane.
Multiferroic order in TbMnO$_{3}$ is analyzed in Ref. \cite{Holbein PRB 23} using polarization described above.

A model of the electromagnon absorption spectra is developed in Ref. \cite{Aupiais npj QM 18} for TbMnO$_{3}$,
where the symmetric type of polarization is considered,
which is basically associated with the electric dipole moment of the following form $\textbf{d}_{12}\sim (\textbf{S}_{1}\cdot\textbf{S}_{2})$.
This model is based on numerical calculations for the Landau-Lifshitz-Gilbert equation,
including a number of terms, including the Dzyaloshinskii-Moriya interaction.
This model includes the detailed description of the location of different atoms/ions in the crystal.
Our model is developed for different regimes of magnetoelectric coupling,
and we consider less detailed structures.
However, our model allows us to get the main experimentally observed feature of the electromagnon phenomenon using the analytical approach.

The generation of the electromagnons in the multiferroic materials via the action of the periodic electric field on the spins
is the experimentally well-known mechanism.
Its possibility is associated with the magnetoelectric coupling.
Usually, this mechanism applied to the noncollinear spin structures.
We have considered the application of this mechanism to the system of collinear spins.
However, we have considered the dynamic of the small-amplitude perturbations creating dynamical noncollinear order,
assuring the appearance of the magneto-electric coupling in this system.
Our model of the imaginary part of the dielectric permeability shows the additional wide peak (electromagnon) in addition to the sharp peak of higher frequency (magnon).
The form and position of the found peaks correspond to the results found in experimental works.
The presented model does not explicitly include the temperature of the system,
so the results have been obtained for temperatures significantly lower than the Curie point.


\emph{DATA AVAILABILITY}:
Data sharing is not applicable to this article as no new data were
created or analyzed in this study, which is a purely theoretical one.


\emph{Acknowledgements}:
The research is supported by the Russian Science Foundation under the
grant
No. 25-22-00064.

\newpage
.
\newpage


\title{Supplementary materials to Mean-field theory of the electromagnon resonance}

\section{Macroscopic spin polarizations of the spin origin}

In order to include the contribution of the electric polarization
in the Landau--Lifshitz--Gilbert equation,
in this paper,
we apply both the well-known macroscopic approach
and the microscopic quantum hydrodynamic method
\cite{AndreevTrukh PS 24}, \cite{AndreevTrukh EPJ B 24}.
We consider the additional term in the energy density of the system
$\Delta \mathcal{E}=-\textbf{P}\cdot\textbf{E}$,
where $\textbf{P}$ is the polarization or the electric dipole moment density,
and $\textbf{E}$ is the electric field.
To consider the evolution of the spin density,
we need to take into account the spin dependence of the polarization
existing in the multiferroics of spin origin.
Before we consider particular mechanisms for the electric dipole formation
and discuss the definition of the polarization.

The macroscopic vector field of polarization is defined
as the quantum average of the sum of
the electric dipole moment operators
acting on the many-particle wave function
(the many-particle wave spinor):
\begin{equation}\label{MFMemf P def}
\textbf{P}(\textbf{r},t)=
\int \Psi_{S}^{\dagger}(R,t)\sum_{i}\delta(\textbf{r}-\textbf{r}_{i})
(\hat{\textbf{d}}_{i}\Psi(R,t))_{S}dR. \end{equation}
The electric dipole moment
operator $\hat{\textbf{d}}_{i}$ is traditionally considered to be the relative shifts of the magnetic ions
from the non-magnetic ion $\hat{\textbf{d}}_{i}=q_{i}\Delta\textbf{r}_{i}$.
It corresponds to the relative shift of ions from positions
that coincide with the center of "mass" for the positive and negative charges.
It can be written via the positions and the dipole moments of the non-magnetic ions.

\subsection{Contribution of polarization in Landau--Lifshitz--Gilbert equation}

If a balance of interparticle interactions forms the noncollinear spin structure in magnetic samples,
it can lead to the further formation of the electric dipole moment due to the shifts of the non-magnetic ions relative to the magnetic ions.
It can be shown that
the spin-orbit interaction and the Heisenberg exchange interaction
forming the magnon spin current lead to the polarization formation.
This polarization corresponds to the following microscopic
electric dipole moment expression
\cite{Tokura RPP 14}, \cite{AndreevTrukh PS 24}
\begin{equation}\label{MFMemf edm def simm}
\textbf{d}_{ij}= \alpha_{ij}[\textbf{r}_{ij}\times[\textbf{s}_{i}\times\textbf{s}_{j}]]. \end{equation}
We make the representation of the equation (\ref{MFMemf edm def simm}) in the operator form associated with the single magnetic ion
\begin{equation}\label{MFMemf edm operator Mod}
\hat{\textbf{d}}_{i}=\sum_{j\neq i}
\alpha_{ij}(r_{ij})[\textbf{r}_{ij}\times[\hat{\textbf{s}}_{i}\times\hat{\textbf{s}}_{j}]], \end{equation}
where the following short-range function is introduced
$\alpha_{ij}(r_{ij})=\alpha_{ij}$
if $r< a_{eff}$,
and
$\alpha_{ij}(r_{ij})=0$ for $r> a_{eff}$.
Following the
quantum hydrodynamic
method
\cite{AndreevTrukh PS 24}, \cite{AndreevTrukh JETP 24}, \cite{AndreevTrukh EPJ B 24},
we can obtain the approximate form of polarization
(\ref{MFMemf P def})
with the electric dipole moment
(\ref{MFMemf edm operator Mod}):
\begin{equation}\label{MFMemf P def expanded} \textbf{P}(\textbf{r},t)=
\sigma
[\textbf{S}(\nabla\cdot \textbf{S})-(\textbf{S}\cdot\nabla)\textbf{S}], \end{equation}
where
$\sigma\equiv -\frac{1}{3}g_{(\alpha)}$,
and
$g_{(\alpha)}=\int \xi^{2}\alpha(\xi) d\mbox{\boldmath $\xi$}$.
This result correspond to
\cite{Mostovoy PRL 06}
and \cite{Dong AinP 15}
(see p. 533).

We consider the contribution of the electric field in the Landau--Lifshitz--Gilbert equation.
It can be found within two methods.
First, we consider the macroscopic method.
Using the macroscopic polarization $\textbf{P}$ (\ref{MFMemf P def expanded}),
we can write the corresponding energy density
$\mathcal{E}$$=-\textbf{P}\cdot\textbf{E}$.
It can also be obtained as
$\langle -\sum_{i}E_{i}^{\beta}\hat{d}_{i}^{\beta}\rangle$
with
$\hat{d}_{i}^{\beta}$ given by the equation (\ref{MFMemf edm operator Mod}).

Let us find the contribution of the interaction of the dipoles with the electric field in
the Landau--Lifshitz--Gilbert equation
\begin{equation}\label{MFMemf LLG eff caused by el f}
\partial_{t}\textbf{S}\mid_{\textbf{E}}=\gamma[\textbf{S}\times\textbf{H}_{eff}], \end{equation}
with
\begin{equation}\label{MFMemf H eff caused by el f}
\textbf{H}_{eff}=-\frac{1}{\gamma}\frac{\delta \mathcal{E}}{\delta \textbf{S}}
=-\frac{1}{\gamma}\biggl(\frac{\partial}{\partial \textbf{S}}
-\partial_{\beta}\frac{\partial}{\partial (\partial_{\beta}\textbf{S})}\biggr)\mathcal{E}. \end{equation}
If we use the equation (\ref{MFMemf P def expanded}) for the polarization in this macroscopic analysis,
we find the following contribution within the
Landau--Lifshitz--Gilbert equation
$$\partial_{t}S^{\alpha}\mid_{\textbf{E}}=
-\sigma
\varepsilon^{\alpha\beta\gamma}S^{\beta}
\biggl[2E^{\mu}\partial^{\gamma}S^{\mu}$$
\begin{equation}\label{MFMemf evol S under infl of E fin 00}
-2E^{\gamma}(\nabla\cdot\textbf{S})
+S^{\mu}\partial^{\gamma}E^{\mu} -(\textbf{S}\cdot\nabla)E^{\gamma}\biggr].\end{equation}
This method is partially based on the microscopic derivation made for the operator (\ref{MFMemf edm operator Mod}).
We apply notation $\sigma\equiv -\frac{1}{3}g_{(\alpha)}=\gamma_{0}\gamma^{2}>0$,
in compare with the parameter $\gamma_{0}$ used in Ref. \cite{Risinggard SR 16}.

The derivation of the polarization contribution in the torque can be completely made using the quantum-hydrodynamic theory
\cite{AndreevTrukh PS 24}, \cite{AndreevTrukh JETP 24}, \cite{AndreevTrukh EPJ B 24}.
Most compact presentation can be made using tensor notations
\begin{equation}\label{MFMemf evol S under infl of E fin 01}
\partial_{t}S^{\alpha}\mid_{\textbf{E}}=\sigma
\varepsilon^{\beta\gamma\delta}\varepsilon^{\delta\mu\nu}\varepsilon^{\alpha\nu\sigma}
[ S^{\mu}S^{\sigma}\partial^{\gamma}E^{\beta} +2E^{\beta}S^{\sigma}\partial^{\gamma}S^{\mu}].
\end{equation}
In order to obtain the equation
(\ref{MFMemf evol S under infl of E fin 00}),
we make the convolution on the index $\delta$ of the first two Levi-Civita symbols.
Another form can be obtained by the convolution on the index $\nu$ of the last two Levi-Civita symbols:
$$\partial_{t}S^{\alpha}\mid_{\textbf{E}}=\sigma
\biggl[S^{2}[\nabla\times\textbf{E}]^{\alpha}
-\varepsilon^{\alpha\beta\gamma}E^{\beta}\partial^{\gamma}S^{2}$$
\begin{equation}\label{MFMemf evol S under infl of E fin}
-S^{\alpha}(\textbf{S}\cdot [\nabla\times\textbf{E}])
+2\varepsilon^{\beta\gamma\delta}S^{\beta}E^{\gamma}\partial^{\delta} S^{\alpha}\biggr].
\end{equation}

\section{Dielectric permeability}

The final form of the Landau--Lifshitz--Gilbert equation has the following form:
$$\partial_{t}\textbf{S}=
A[\textbf{S}\times\triangle\textbf{S}]
+\kappa [\textbf{S}\times S_{z}\textbf{e}_{z}] $$
$$-\sigma
\biggl[ [\textbf{E}\times \nabla] S^{2}
-2(\textbf{S}\cdot[\textbf{E}\times\nabla]) \textbf{S}$$
\begin{equation}\label{MFMemf s evolution}
-S^{2}(\nabla\times\textbf{E})
+\textbf{S}(\textbf{S}\cdot [\nabla\times\textbf{E}])
\biggr]
+a[\textbf{S}\times\partial_{t}\textbf{S}]. \end{equation}
The equation (\ref{MFMemf s evolution}) presents the spin density evolution equation obtained in the mean-field approximation,
with the phenomenologically included damping in the Gilbert form (the last term, proportional to constant $a<0$).
Quasi-classic and quantum spin currents associated with the displacement of particles (atoms or ions) are neglected.
The right-hand side contains three groups of terms in addition to the Gilbert damping.
The long-wavelength limit of the exchange interaction is presented by the first term proportional to the constant $A>0$.
The contribution of the anisotropy energy associated with the anisotropy of the Heisenberg exchange interaction Hamiltonian,
which is also considered in the long-wavelength limit is presented with the second term proportional to the constant $\kappa>0$ \cite{KOSEVICH PR 90}.
The third group of terms proportional to $\sigma>0$ presents the magnetoelectric coupling described in the
previous section.

\subsection{Periodic electric field, Stationary regime and Perturbations}

We consider the system of parallel spins as the stationary regime
$\textbf{S}_{0}=S_{0}\textbf{e}_{z}$.
We place this system in the periodic electric field
$\textbf{E}_{0}=\epsilon_{0}\textbf{e}_{y}\cos(\omega_{0} t-k_{0}x)$
as the part of the stationary regime ($\omega_{0}=k_{0}c$).
The equation (\ref{MFMemf s evolution}) demonstrates a possibility of the stationary state in the chosen regime.
Thus, we have the zero value for the
right-hand side of equation (\ref{MFMemf s evolution}).
Further, we consider the perturbations of the found stationary state,
and calculate the evolution of perturbations in both the spin density and the electric field.
To include the electric field perturbations,
we apply the Maxwell's equations
\begin{equation}\label{MFMemf } \nabla\times \textbf{E}=-\frac{1}{c}\partial_{t}\textbf{B},\end{equation}
and
\begin{equation}\label{MFMemf }
\nabla\times \textbf{B}=\frac{1}{c}\partial_{t}\textbf{E} +4\pi\nabla\times \textbf{M}+\frac{4\pi}{c}\partial_{t}\textbf{P}, \end{equation}
where we represent the magnetization via the spin density
$\textbf{M}=\gamma\textbf{S}$ and include the gyromagnetic ratio $\gamma$.

We can keep $\partial_{t}\textbf{E}$ in order to get the electromagnetic wave dispersion dependence existing in the medium along with the spin wave.
This contribution can be dropped at the analysis of the general dispersion equation.
The term $\partial_{t}\textbf{P}$ gives a correction to $\partial_{t}\textbf{E}$,
so we drop it in comparison with the main term.

Both the electromagnetic waves and the spin waves can propagate in the dielectric magnetically ordered medium
(we do not include the acoustic waves in this paper).
In addition to the electromagnetic eigen-modes,
we pass the external harmonic electromagnetic wave through the infinite sample.
No constant electric or magnetic field is applied to the sample.
Moreover, we do not include the Zeeman energy
that describes the action of the magnetic part of the wave on the magnetic moments of the medium.
The dipole-dipole interaction of the magnetic moments is excluded as well.

Basically, the evolution of the small amplitude perturbations in the magnetically ordered dielectric materials
under the action of the periodic electric field of the external electromagnetic wave is considered.
The collinear order of spins is chosen as the part of the stationary state.
The directions of the equilibrium spin density, the external wave propagation direction,
and the amplitude of the external electric field form the right vector triple.
The magnetoelectric coupling associated with the noncollinear spin order affects the dynamics of the system.
It gives the nonzero value of the spin-torque at the propagation of the spin density perturbations.
Dissipation in the stable state corresponds to the positive imaginary part of the dielectric
susceptibility for the chosen structure of the perturbations.
The magnetoelectric coupling
leads to the negative value of
the external frequency dependent contribution to
the imaginary part of the dielectric
susceptibility $Im\varepsilon_{zz,fd}$.
However, the frequency dependent part of the susceptibility $Im\varepsilon_{zz,fd}$ appears
as the relativistic correction to the positive main part of $Im\varepsilon_{zz,mp}$
which contains no dependence on the external wave frequency.

\subsection{Relations between harmonics of the magnetic field, electric field and spin density}

Let us present the linearized form of equation (\ref{MFMemf s evolution})
$$\partial_{t}\delta\textbf{S}=
A[\textbf{S}_{0}\times\triangle\delta\textbf{S}]
-\kappa [\textbf{S}_{0}\times \delta\textbf{S}] +a[\textbf{S}\times\partial_{t}\delta\textbf{S}]$$
\begin{equation}\label{MFMemf s evolution lin}
+\sigma
\biggl[ S_{0}^{2}(\textbf{e}_{x}\times\partial_{x}\delta\textbf{E})
-2S_{0} \epsilon_{0}\cos(\omega_{0} t-k_{0}x)\partial_{x}\delta\textbf{S}\biggr]
. \end{equation}
The last term shows that
we deal with the
partial differential equation set with
the changing coefficient.
Its solution can be found in the form of the plane waves,
but we need to consider harmonics of the frequency and wave vector of the external wave.

We focus on the perturbations
that propagate perpendicular to the stationary spin density $\textbf{S}_{0}=S_{0}\textbf{e}_{z}$,
in the direction parallel to the external electromagnetic wave propagation direction
\begin{equation}\label{MFMemf } \delta\textbf{S}=\sum_{n=-\infty}^{+\infty}S_{n}^{\alpha}e^{-\imath\omega t+\imath kx-\imath\omega_{0}nt+\imath k_{0}nx},\end{equation}
where we include the harmonics of the external field.
The small value of the terms containing the external electric filed allows us to consider $n\in [-1,1]$:
\begin{equation}\label{MFMemf }
\delta\textbf{S}=\sum_{n=-1}^{+1}S_{n}^{\alpha}e^{-\imath\omega t+\imath kx-\imath\omega_{0}nt+\imath k_{0}nx}\end{equation}
due to the small value of constant $\sigma$ and, if it is necessary, the small amplitude $\epsilon_{0}$.

This leads
(in the combination with the linearized form of the Maxwell equations)
to the simplified form for the electric and magnetic fields generated in the system
\begin{equation}\label{MFMemf } \delta\textbf{B}=\sum_{n=-1}^{+1}B_{n}^{\alpha}e^{-\imath\omega t+\imath kx-\imath\omega_{0}nt+\imath k_{0}nx},\end{equation}
and
\begin{equation}\label{MFMemf } \delta\textbf{E}=\sum_{n=-1}^{+1}E_{n}^{\alpha}e^{-\imath\omega t+\imath kx-\imath\omega_{0}nt+\imath k_{0}nx}.\end{equation}

The linearized Landau--Lifshitz--Gilbert equation (\ref{MFMemf s evolution lin})
allows us to express the perturbations of the spin density
\begin{equation}\label{MFMemf }\hat{\xi}_{S,6}\equiv\left(
                   \begin{array}{c}
                     S_{x,-1} \\
                     S_{y,-1}  \\
                     S_{x,0}  \\
                     S_{y,0}  \\
                     S_{x,1}  \\
                     S_{y,1}  \\
                   \end{array}
                 \right),
\end{equation}
via the perturbations of the electric field
\begin{equation}\label{MFMemf } \hat{\xi}_{E,6}\equiv\left(
                                  \begin{array}{c}
                                    0 \\
                                    \sigma S_{0}^{2} (k-k_{0})E_{z,-1} \\
                                    0 \\
                                    \sigma S_{0}^{2} k E_{z,0} \\
                                    0 \\
                                    \sigma S_{0}^{2} (k+k_{0})E_{z,1} \\
                                  \end{array}
                                \right)
\end{equation}
in the following form
\begin{equation}\label{MFMemf matrix equation short} \hat{\Delta}_{E,6\times6}\hat{\xi}_{S,6}=\hat{\xi}_{E,6},\end{equation}
where
\begin{widetext}
\begin{equation}\label{MFMemf } \hat{\Delta}_{E,6\times6}\equiv\left(
                                  \begin{array}{cccccc}
                                    \omega-\omega_{0} & -\imath \Omega_{-} & -\sigma S_{0}\epsilon_{0}(k-\frac{1}{2}k_{0}) & 0 & 0 & 0 \\
                                    \imath \Omega_{-} & \omega-\omega_{0} & 0 & -\sigma S_{0}\epsilon_{0}k & 0 & 0  \\
                                    -\sigma S_{0}\epsilon_{0}(k-\frac{1}{2}k_{0}) & 0 & \omega & -\imath \Omega_{0} & -\sigma S_{0}\epsilon_{0}(k+\frac{1}{2}k_{0}) & 0 \\
                                    0 & -\sigma S_{0}\epsilon_{0}(k-k_{0}) & \imath \Omega_{0} & \omega & 0 & -\sigma S_{0}\epsilon_{0}(k+k_{0})  \\
                                    0 & 0 & -\sigma S_{0}\epsilon_{0}(k+\frac{1}{2}k_{0}) & 0 & \omega+\omega_{0} & -\imath \Omega_{+}  \\
                                    0 & 0 & 0 & -\sigma S_{0}\epsilon_{0}k & \imath \Omega_{+} & \omega+\omega_{0}  \\
                                  \end{array}
                                \right),
\end{equation}
where
$\Omega_{0}\equiv (\kappa+\imath \omega a+Ak^{2})S_{0}$,
$\Omega_{+}\equiv (\kappa+\imath(\omega+\omega_{0})a+A(k+k_{0})^{2})S_{0}$,
and
$\Omega_{-}\equiv (\kappa+\imath(\omega-\omega_{0})a+A(k-k_{0})^{2})S_{0}$.

\emph{Spin solution}:
Analysis of equation
(\ref{MFMemf matrix equation short})
leads to the spin solution for the zero harmonic
\begin{equation}\label{MFMemf Sx via B D}
\delta S_{x,n=0}=\imath\frac{k }{\Delta_{E}}\sigma S_{0}^{2}\delta E_{z,0}
\biggl\{\Omega_{0}\mathcal{B}^{2}\mathcal{D}^{2}
 +sk\biggl[\mathcal{B}^{2}\Omega_{-}s\biggl(k-\frac{k_{0}}{2}\biggr)
+\mathcal{D}^{2}s\biggl(k-\frac{k_{0}}{2}\biggr)\Omega_{+}\biggr]\biggr\},
\end{equation}
and
\begin{equation}\label{MFMemf Sy via B D}
\delta S_{y,n=0}=\frac{k }{\Delta_{E}}\sigma S_{0}^{2}\delta E_{z,0}
\biggl\{
\omega
\mathcal{B}^{2}\mathcal{D}^{2}
+\mathcal{B}^{2} \biggl(s\biggl(k-\frac{k_{0}}{2}\biggr)\biggr)^{2}(\omega-\omega_{0})l
-\mathcal{D}^{2}\biggl(s\biggl(k+\frac{k_{0}}{2}\biggr)\biggr)^{2}(\omega+\omega_{0})j\biggr\}
\end{equation}
as the function of the zero harmonic perturbations of the electric field $\delta E_{z,0}$.
Which allows to get the dielectric permeability of the system.
Solutions (\ref{MFMemf Sx via B D}) and (\ref{MFMemf Sy via B D}) include

$$\Delta_{E}= \mathcal{B}^{2}\mathcal{D}^{2} (\omega^{2}-\Omega_{0}^{2})$$
$$-\mathcal{D}^{2}\biggl(s\biggl(k+\frac{k_{0}}{2}\biggr)\biggr)^{2}[\omega(\omega+\omega_{0})j-s^{2}k(k+k_{0})+2\Omega_{0}\Omega_{+}]
-\mathcal{D}^{2}s^{2}k(k+k_{0}) \omega(\omega+\omega_{0})$$
$$-\mathcal{B}^{2}\biggl(s\biggl(k-\frac{k_{0}}{2}\biggr)\biggr)^{2}[\omega(\omega-\omega_{0})l-s^{2}k(k-k_{0})+2\Omega_{0}\Omega_{-}]
-\mathcal{B}^{2}s^{2}k(k-k_{0}) \omega(\omega-\omega_{0})$$
\begin{equation}\label{MFMemf Delta E 01}
+(\omega^{2}-\omega_{0}^2)sk \biggl[j s(k-k_{0})\biggl(s\biggl(k+\frac{k_{0}}{2}\biggr)\biggr)^{2}
+l s(k+k_{0})\biggl(s\biggl(k-\frac{k_{0}}{2}\biggr)\biggr)^{2}\biggr]
-2\Omega_{+}\Omega_{-}(sk)^{2}s^{2}\biggl(k^{2}-\frac{k_{0}^{2}}{4}\biggr) \end{equation}
\end{widetext}
and several additional notations
$s\equiv \sigma S_{0}\epsilon_{0}$,
$\mathcal{B}^{2}\equiv (\omega+\omega_{0})^{2}j-\Omega_{+}^{2}$,
and
$\mathcal{D}^{2}\equiv (\omega-\omega_{0})^{2}l-\Omega_{-}^{2}$.

Our calculation includes the relations between the spin density, the electric field and the magnetic field perturbations
\begin{equation}\label{MFMemf }
\delta B_{y,\pm1}=
4\pi\gamma\frac{(k\pm k_{0})^{2}}{(k-\omega/c)(k\pm2k_{0}+\omega/c)}\delta S_{y,\pm1},
\end{equation}
and
\begin{equation}\label{MFMemf }
\delta E_{z,\pm1}= -\frac{\omega\pm\omega_{0}}{(k\pm k_{0})c}\delta B_{y,\pm1}.
\end{equation}

Relation between $\delta B_{y,\pm1}$ and $\delta S_{y,\pm1}$ leads to the following notation
\begin{equation}\label{MFMemf }
\eta_{B,\pm}\equiv 4\pi\gamma^{2}S_{0} \frac{1}{(k-\omega/c)(k\pm2k_{0}+\omega/c)}.
\end{equation}
We also introduce two parameters:
$j\equiv 1+\sigma S_{0}\eta_{B,+}(k+k_{0})^2 /\gamma c$,
and
$l\equiv 1+\sigma S_{0}\eta_{B,-}(k-k_{0})^2 /\gamma c$.

In this regime,
we get perturbations of the polarization in one direction parallel to the equilibrium spin density
\begin{equation}\label{MFMemf P via Sx}
\delta P_{z,n=0}
=\sigma S_{0}\imath k\delta S_{x,0}/\gamma\mu.
\end{equation}
Hence, we find one nontrivial component of the dielectric permeability tensor
$\delta P_{z,n=0}
\equiv
\kappa_{zz}
\delta E_{z,n=0}
\equiv
(\varepsilon_{zz}-1)
\delta E_{z,n=0}/4\pi$.

\subsection{Dispersion dependence}

In order to continue our analysis,
we find the dispersion dependence.

We present the dispersion dependence for the "low-frequency" excitations $\omega\ll kc$:
$$\Delta_{E}=
-4\pi\gamma
\frac{\omega}{c}\sigma S_{0}^{2}
\biggl\{
\omega
\mathcal{B}^{2}\mathcal{D}^{2}$$
\begin{equation}\label{MFMemf dispersion equation via Delta}
+\mathcal{B}^{2} \biggl(s\biggl(k-\frac{k_{0}}{2}\biggr)\biggr)^{2}(\omega-\omega_{0})l
-\mathcal{D}^{2}\biggl(s\biggl(k+\frac{k_{0}}{2}\biggr)\biggr)^{2}(\omega+\omega_{0})j\biggr\},
\end{equation}
where
$\Delta_{E}$ is given by equation (\ref{MFMemf Delta E 01}).

Both the spin solution (\ref{MFMemf Sx via B D}) and (\ref{MFMemf Sy via B D})
and the dispersion equation (\ref{MFMemf dispersion equation via Delta})
have a complex structure,
and contain different degrees of the relativistic terms.
While the relativistic effects can be small.
So, the found expressions can be expanded to specify
the hierarchy of the relativistic effects.

Let us present the expression for the spin density perturbation in the lowest order on the relativistic corrections
\begin{equation}\label{MFMemf }
\delta S_{x,n=0}=
\sigma S_{0}^{2}\delta E_{z,0} \frac{k  \imath\Omega_{0}}{(\omega^{2}-\Omega_{0}^{2})},
\end{equation}
and
\begin{equation}\label{MFMemf }
\delta S_{y,n=0}=
\sigma S_{0}^{2}\delta E_{z,0} \frac{k  \omega}{(\omega^{2}-\Omega_{0}^{2})}.
\end{equation}
It corresponds to dropping terms proportional to $s\equiv \sigma S_{0}\epsilon_{0}$.
However, the non-zero value of the spin perturbations $\delta \textbf{S}_{n=0}$
as a function of the electric field $\delta E_{z,0}$
(showing the dielectric properties of the material)
is proportional to the coefficient $\sigma$,
which describs the contribution of polarization.

\subsubsection{Simplified dispersion equation}

On the next step, we consider a generalization of the simplest form of the spin density,
but we focus on the narrow interval of frequencies of the external wave.
The lowest order of relativistic terms in the dispersion equation
(\ref{MFMemf dispersion equation via Delta})
gives
$\omega\approx\Omega=(\kappa+Ak^2)S_0$
(if we include no dissipation).
We are also interested in $\omega_{0}\leq\omega\approx\Omega$,
which approximately corresponds to $k_0\ll k$.
We also drop terms proportional to $s^{4}$ in compare with terms of the lower order of $s$
in
$$\eta_{B,-}=\eta_{B,+}=\frac{4\pi\gamma^{2}S_{0}}{k^{2}}$$
with
$j \rightarrow 1+(\sigma S_{0}/\gamma c)\eta_{B,+} k^2$,
$l \rightarrow 1+(\sigma S_{0}/\gamma c)\eta_{B,-} k^2$,
$j=l= 1+\alpha$,
where $\alpha=4\pi\gamma\sigma S_{0}^{2}/c$
and equation (\ref{MFMemf Delta E 01}) for determinant $\Delta_{E}$ simplifies to
$$\Delta_{E}=\mathcal{B}^{2}\mathcal{D}^{2}(\omega^{2}-\Omega_{0}^{2})$$
$$-(sk)^{2}\biggl[\mathcal{D}^{2}[\omega(\omega+\omega_{0})(1+\alpha) +2\Omega_{0}\Omega_{+}]
-\mathcal{D}^{2} \omega(\omega+\omega_{0})$$
\begin{equation}\label{MFMemf Delta E 02}
-\mathcal{B}^{2}[\omega(\omega-\omega_{0})(1+\alpha) +2\Omega_{0}\Omega_{-}]
-\mathcal{B}^{2} \omega(\omega-\omega_{0}) \biggr].
\end{equation}

This also provides the simplified dispersion dependence corresponding to
(\ref{MFMemf Delta E 02}):
$$\Delta_{E}=
-\omega \alpha
\biggl\{
\omega
\mathcal{B}^{2}\mathcal{D}^{2}$$
\begin{equation}\label{MFMemf dispersion equation via Delta no k0}
+(sk)^{2}[\mathcal{B}^{2} (\omega-\omega_{0})(1+\alpha)
-\mathcal{D}^{2}(\omega+\omega_{0})(1+\alpha)]\biggr\}
.\end{equation}
Let us repeat here few modified notations
notations
$s\equiv \sigma S_{0}\epsilon_{0}$,
$\mathcal{B}^{2}\equiv (\omega+\omega_{0})^{2}(1+\alpha)-\Omega_{+}^{2}$,
and
$\mathcal{D}^{2}\equiv (\omega-\omega_{0})^{2}(1+\alpha)-\Omega_{-}^{2}$,
with
$\Omega_{+}= (\kappa+\imath(\omega+\omega_{0})a+Ak^{2})S_{0}$,
and
$\Omega_{-}= (\kappa+\imath(\omega-\omega_{0})a+Ak^{2})S_{0}$.

In our analysis,
we need to use the frequency obtained in the zero order on $\sigma$
($\alpha\sim\sigma$ and $s\sim\sigma$)
\begin{equation}\label{MFMemf dispersion equation via Delta s to 0}
\mathcal{B}_{0}^{2}\mathcal{D}_{0}^{2}(\omega^{2}-\Omega_{0}^{2})=0,
\end{equation}
where
we introduce
$\mathcal{B}_{0}^{2}\equiv (\omega+\omega_{0})^{2}-\Omega_{+}^{2}$,
and
$\mathcal{D}_{0}^{2}\equiv (\omega-\omega_{0})^{2}-\Omega_{-}^{2}$.
It gives
\begin{equation}\label{MFMemf dispersion omega 0 order}
\omega^{2}=\Omega_{0}^{2}
\end{equation}
leading to
$\omega=\nu\Omega(1+\imath \tilde{a})$,

\subsubsection{Simplified dielectric permeability}

In order to obtain
the dielectric susceptibility
(in the described approximation with $s$-contribution up to $s^{2}$, but no $k_0$ contribution)
we replace $\Delta_{E}$ in the spin density solution (\ref{MFMemf Sx via B D}) required for the polarization
(\ref{MFMemf P via Sx})
via the dispersion equation (\ref{MFMemf dispersion equation via Delta no k0})
$$\kappa_{zz}
= \sigma S_{0}\frac{k^{2}c}{4\pi\gamma^{2}\mu\omega}$$
\begin{equation}\label{MFMemf kappa zz fin}\times
\frac{\biggl\{\Omega_{0}\mathcal{B}^{2}\mathcal{D}^{2}
+(sk)^2 \biggl[\mathcal{B}_{0}^{2}\Omega_{-}
+\mathcal{D}_{0}^{2}\Omega_{+}\biggr]\biggr\}}
{ \biggl\{
\omega
\mathcal{B}^{2}\mathcal{D}^{2}
+(sk)^2[\mathcal{B}_{0}^{2} (\omega-\omega_{0})
-\mathcal{D}_{0}^{2}(\omega+\omega_{0})] \biggr\}}
\end{equation}
where
we use
$\mathcal{B}_{0}^{2}\equiv (\omega+\omega_{0})^{2}-\Omega_{+}^{2}$,
and
$\mathcal{D}_{0}^{2}\equiv (\omega-\omega_{0})^{2}-\Omega_{-}^{2}$.

Dropping the relativistic corrections to the main terms,
we obtain
$\kappa_{zz}=\frac{\sigma S_{0}}{\gamma\mu}k^{2}c
\frac{\Omega_{0}}{4\pi\gamma\omega^{2}}$
and find no resonances here
for the dielectric susceptibility as the function of the external wave frequency.

\section{Structure of the dielectric susceptibility}

Equation (\ref{MFMemf kappa zz fin}) shows the final form of the dielectric susceptibility obtained up to $\alpha^{2}$ approximation.
Generally, it requires the frequency obtained from equation (\ref{MFMemf dispersion equation via Delta no k0})
up to $\alpha^{2}$ in terms of the zero order on $\alpha$.
But we need zero order approximation on $\alpha$ the frequency (\ref{MFMemf dispersion omega 0 order}) in terms proportional to $\alpha^{2}$.

The dielectric susceptibility can be split into two parts
$\kappa_{zz}=\kappa_{zz,main}+\kappa_{zz,rel}$,
where
$\kappa_{zz,main}$ is the part with no explicit dependence on $\alpha$,
but it contains dependence on $\alpha$ via the frequency,
and
$\kappa_{zz,rel}$ is the part proportional to $\alpha^{2}$.

As it is mentioned in the text,
the first part of the dielectric susceptibility
\begin{equation}\label{MFMemf kappa zz main 0 on alpha}
\kappa_{zz,main}
=\kappa_{0}\frac{\Omega}{\omega}\frac{\Omega_{0}}{\omega}
\approx\kappa_{0}\frac{\Omega}{\Omega_{0}}
, \end{equation}
with
\begin{equation}\label{MFMemf}
\kappa_{0}=\frac{\sigma\hbar S_{0}k^2 c}{4\pi\mu\Omega}
\end{equation}
shows no resonances.
It contains some non-resonance dependence on $\omega_{0}$ in the part
where we consider the frequency $\omega(k)$ up to $\alpha^{2}$.
The imaginary part of the first part of the dielectric susceptibility is positive (\ref{MFMemf kappa zz main 0 on alpha}).

Let us present the simplified form of equation (\ref{MFMemf kappa zz fin})
$$\kappa_{zz}=\sigma S_{0}\frac{k^{2}c\Omega_{0}}{4\pi\gamma^{2}\mu\omega^{2}}\times
\biggl\{1
+(sk)^2 \frac{[\mathcal{B}_{0}^{2}\Omega_{-}
+\mathcal{D}_{0}^{2}\Omega_{+}]}{\Omega_{0}\mathcal{B}_{0}^{2}\mathcal{D}_{0}^{2}}$$
\begin{equation}\label{MFMemf kappa zz 2}
-(sk)^2\frac{[\mathcal{B}_{0}^{2} (\omega-\omega_{0})
-\mathcal{D}_{0}^{2}(\omega+\omega_{0})]}{\omega
\mathcal{B}_{0}^{2}\mathcal{D}_{0}^{2}}
\biggr\}.
\end{equation}
It can be split into the real and imaginary parts
$\kappa_{zz}=Re\kappa_{zz}+\imath Im\kappa_{zz}$,
while we further focus on
the imaginary part of the dielectric susceptibility as the function of the dimensionless frequency of the external wave:
\begin{equation}\label{MFMemf}
Im\kappa_{zz,rel}=\kappa_{0}\frac{(sk)^{2}}{\Omega^{2}}(1+\tilde{a}^{2})^{3}\Xi
, \end{equation}
where
\begin{widetext}
$$\Xi=
\Biggl\{
2[\nu^{4}(1-\tilde{a}^{2})+x^3-x\nu^{2}(1+\nu)+x^2\nu(1+\tilde{a}^2\nu)]$$
$$\cdot\Biggl[
\tilde{a}(1-3\tilde{a}^2)\biggl([4\tilde{a}^2\nu^{2}-x(1-\nu)][x^2(1+\tilde{a}^2)+2x-\tilde{a}^2\nu^{2}]
+[4\tilde{a}^2\nu^{2}+x(1-\nu)][x^2(1+\tilde{a}^2)-2x-\tilde{a}^2\nu^{2}]\biggr)$$
$$-\tilde{a}(3-\tilde{a}^2)\biggl([x^2(1+\tilde{a}^2)+2x-\tilde{a}^2\nu^{2}][x^2(1+\tilde{a}^2)-2x-\tilde{a}^2\nu^{2}]
-\tilde{a}^2[16\tilde{a}^4\nu^{4}-x^2(1-\nu)^2]\biggr)\Biggr]$$
$$+2\tilde{a}\nu [2\nu^{3}+x^3 (1+\tilde{a}^2)-x^2 \nu \tilde{a}^4-x\nu(3+\tilde{a}^2+\tilde{a}^4 \nu)] $$
$$\cdot\Biggl[ (1-3\tilde{a}^2)
\biggl([x^2(1+\tilde{a}^2)+2x-\tilde{a}^2\nu^{2}][x^2(1+\tilde{a}^2)-2x-\tilde{a}^2\nu^{2}]
-\tilde{a}^2[16\tilde{a}^4\nu^{4}-x^2(1-\nu)^2]\biggr)$$
$$+\tilde{a}^2(3-\tilde{a}^2)\biggl([4\tilde{a}^2\nu^{2}-x(1-\nu)][x^2(1+\tilde{a}^2)+2x-\tilde{a}^2\nu^{2}]
+[4\tilde{a}^2\nu^{2}+x(1-\nu)][x^2(1+\tilde{a}^2)-2x-\tilde{a}^2\nu^{2}]\biggr)
\Biggr]\Biggr\}$$
$$\cdot\Biggl[(1-3\tilde{a}^2)^2+\tilde{a}^2 (3-\tilde{a}^2)^2\Biggr]^{-1}
\cdot\Biggl[[x^2(1+\tilde{a}^2)+2x-\tilde{a}^2\nu^{2}]^2+\tilde{a}^2[4\tilde{a}^2\nu^2+x(1-\nu)]^2\Biggr]^{-1}$$
\begin{equation}\label{MFMemf Xi explicit}
\cdot\Biggl[[x^2(1+\tilde{a}^2)-2x-\tilde{a}^2\nu^{2}]^2+\tilde{a}^2[4\tilde{a}^2\nu^2-x(1-\nu)]^2\Biggr]^{-1},
\end{equation}
\end{widetext}
where
$x=\omega_{0}/\Omega$,
and
$\nu=(1+\tilde{a}^2)^{-1}$.

\end{document}